\documentclass[11pt]{article}
\pdfoutput=1

\usepackage{stolenstyle}

\usepackage{amsmath,epsf,amssymb,latexsym,amsthm,setspace,bbm,array,pifont,enumerate}

\DeclareFontFamily{U}{rsf}{}
\DeclareFontShape{U}{rsf}{m}{n}{
  <5> <6> rsfs5 <7> <8> <9> rsfs7 <10-> rsfs10}{}
\DeclareMathAlphabet\Scr{U}{rsf}{m}{n}

\def\CO#1#2{{[#1,#2]}}

\def\iden{{\mathbbm 1}}

\def\GUL{\GU(1)_{\text{L}}}
\def\GUR{\GU(1)_{\text{R}}}

\def\C{{\mathbb C}}

\def\P{{\mathbb P}}
\def\R{{\mathbb R}}
\def\Z{{\mathbb Z}}

\def\Hom{\operatorname{Hom}}

\def\Rea{\operatorname{Re}}

\def\SO{\operatorname{SO}}
\def\SL{\operatorname{SL}}
\def\GL{\operatorname{GL}}

\def\GO{\operatorname{O{}}}
\def\SU{\operatorname{SU}}
\def\GU{\operatorname{U{}}}

\def\Spin{\operatorname{Spin}}
\def\GG{\operatorname{G}}

\def\spin{\operatorname{\mathfrak{spin}}}

\def\la{\langle}
\def\ra{\rangle}

\def\ff#1#2{{\textstyle\frac{#1}{#2}}}

\def\cC{{\cal C}}

\def\cF{{\cal F}}

\def\cM{{\cal M}}
\def\cN{{\cal N}}

\def\cP{{\cal P}}

\def\cR{{\cal R}}
\def\cS{{\cal S}}
\def\cT{{\cal T}}

\def\cX{{\cal X}}

\def\cZ{{\cal Z}}

\newcommand\kappah{\widehat{\kappa}}

\newcommand\sigmah{\widehat{\sigma}}

\newcommand\Omegab{\overline{\Omega}}

\newcommand\gh{\widehat{g}}

\newcommand\Gh{\widehat{G}}

\newcommand\Mb{\overline{M}}

\newcommand\Xb{\overline{X}}
\newcommand\Yb{\overline{Y}}
\newcommand\Zb{\overline{Z}}

\theoremstyle{definition}

\usepackage{upgreek}

\usepackage[dvipsnames]{xcolor}

\usepackage{tikz}

\usetikzlibrary[shapes.geometric]
\usetikzlibrary{positioning} 
\usetikzlibrary{calc,intersections,through,backgrounds}
\usetikzlibrary{cd}

\tikzset{>=stealth}
\tikzset{every picture/.style={very thick}}

\def\ii{\mathrm{i}}
\def\ed{\mathrm{d}}

\def\bF{{\boldsymbol{F}}}

\def\sleft{\text{{\tiny{L}}}}
\def\sright{\text{\tiny{R}}}

\def\NSNS{{{\text{\tiny{NS-NS}}}}}
\def\RNS{{{\text{\tiny{R-NS}}}}}
\def\NSR{{{\text{\tiny{NS-R}}}}}
\def\RR{{{\text{\tiny{R-R}}}}}

\def\Hol{{\operatorname{Hol}}}

\def\preprint{ ~~ }

\title{Unorientable supersymmetric compactifications of M-theory}
\author[a] {Ilarion V.~Melnikov}
\author[b] {and Ruben Minasian}
\affiliation[a] {Department of Physics and Astronomy,
James Madison University,
Harrisonburg, VA 22807, USA}
\affiliation[b] {Institut de Physique Th{\'e}orique,  
Universit{\'e} Paris-Saclay, CNRS, CEA, F-9119, Gif-sur-Yvette, France
}

\emailAdd{melnikix@jmu.edu}
\emailAdd{ruben.minasian@ipht.fr}

\abstract{We discuss orbifold compactifications of the type II string, where the orbifold combines a free action on a base manifold and the $(-1)^{\bF_{\sleft}}$ symmetry which acts by $-1$ on left-moving Ramond sectors.  Such backgrounds can preserve minimal supersymmetry and lift to M-theory compactification on smooth unorientable manifolds.  }

\begin{document}

\maketitle

\section{Introduction} \label{s:Introduction}

 $(-1)^{\bF_{\sleft}}$ is a symmetry in type II string theory that reverses the sign of states in the left moving Ramond sector. The orbifold of the IIA string by $(-1)^{\bF_{\sleft}}$ yields the IIB string and vice versa~\cite{Vafa:1995gm}.   If the ten-dimensional background is a compactification $\R^{1,d-1} \times X$, the internal manifold $X$ may admit a free action by some group $G$. $X$ and the quotient manifold $Y=X/G$ generally have different topologies, and $Y$ may preserve some fraction (possibly all) of the supersymmetry preserved by $X$. The present study is motivated by two questions:  When can these two different actions be combined in a non-trivial yet consistent fashion?  What is the resulting orbifold theory?

There is one well-studied example of combining these two actions---when $X$ is taken to be a circle, and $G=\mathbb{Z}_2$ acts by a half-period shift. In this special case $X$ and $Y$ are both circles, with the radius of $Y$ half that of $X$, and they clearly preserve the same amount of supersymmetry. The compactification on $X=S^1$ to nine dimensions with the quotient by the combined action of the two symmetries yields a theory with 16 supercharges - a minimal supergravity coupled to a single vector multiplet~\cite{Dabholkar:1996pc, Aharony:2007du}. Such a compactification is possible both for IIA and IIB, and the resulting $9$-dimensional theories are the same only at the massless perturbative level; further circle reduction yields the same eight-dimensional theory. Interestingly these theories are qualitatively different from other consistent nine-dimensional theories with the same amount of supersymmetry, both in terms of its symmetries and phase structure~\cite{Hellerman:2005ja,Aharony:2007du,Montero:2022vva}, and the structure of higher derivative couplings~\cite{Bossard:2026pex}. The $(-1)^{\bF_{\sleft}}$ action in IIA  lifts to a reflection of the M-theory circle, meaning that the IIA version of this quotient can be lifted to eleven dimensions and geometrized: it is equivalent to the compactification of M-theory on a Klein bottle K.

It is believed that M-theory can be consistently compactified on manifolds with a $\text{pin}^+$ structure.   The evidence for this originates in the lifts of seemingly well-behaved type II backgrounds~\cite{Vafa:1995gm}, as well studies of the parity symmetry on an M2 brane~\cite{Witten:2016cio} and in $11$--dimensional supergravity~\cite{Witten:1996md, Freed:2019sco}.   Since unorientable manifolds are not spin, the possibility of preserving supersymmetry might seem surprising, yet it can be realized and, moreover, without the introduction of orientifold planes or intrinsically strongly-coupled effects familiar in typical F-theory compactifications.
So far the Klein bottle and a three-dimensional Bieberbach manifold (yielding an eight-dimensional theory which at the perturbative level is the same as  $\text{K}\times S^1$ reduction of M-theory) seem to be the only unorientable supersymmetric compactifications discussed in the literature.\footnote{In context of AdS compactifications, reduction on $\mathbb{R}\mathbb{P}^4$ to maximally supersymmetric theory on $\text{AdS}_7$ was discussed in~\cite{Bonetti:2024etn}.} In this note we will make some observations on supersymmetric backgrounds on unorientable manifolds, and discuss four-dimensional Minkowski theories with ${\mathcal N}=1$ supersymmetry arising in such compactifications.

Our main example is already mentioned in~\cite{Vafa:1995gm}:  an orbifold of the IIA string compactified on a smooth Calabi-Yau (CY) space $X$ by an involution $g(-1)^{\bF_{\sleft}}$, where $g$ is a generator for a freely-acting K\"ahler $G=\Z_2$ isometry of $X$.\footnote{Since~\cite{Vafa:1995gm} many works have considered quotients involving a combination of $(-1)^{\bF_{\sleft}}$ with a geometric, or, more generally, some asymmetric orbifold action, for example~\cite{Sen:1996tz,Angelantonj:1996uy,Kachru:2001je}.  Unlike those studies, our focus is on the situation where the combined action has a geometric realization and is free.}  This can be interpreted geometrically as a compactification on the non-simply connected manifold $Y=X/G$ with a holonomy for $(-1)^{\bF_{\sleft}}$ supported by non-contractible cycles of $Y$.  Another geometric interpretation is through an M-theory lift to an orbifold of $X\times S^1$ by an involution $\kappa$, which acts as $g$ on $X$ combined with a reflection on the circle coordinate.  

These examples deserve further study for a number of reasons.  First, as already pointed out in~\cite{Vafa:1995gm}, while preserving only $4$ supercharges, they only involve smooth geometry without orientifold planes or branes, and the two-derivative string tree-level effective action for the massless fields is determined by the (2,2) superconformal field theory on the world-sheet and can be computed exactly.\footnote{This can be contrasted with the so-called ``barely $\GG_2$'' manifolds discussed in~\cite{Joyce:1996crm,Harvey:1999as,Grigorian:2009nx}, where only $(1,1)$ worldsheet supersymmetry is preserved by the quotient. These oriented seven-manifolds of holonomy $SU(3) \rtimes \Z_2$ will be revisited in Section \ref{sec:discussion}.}  Indeed, the quotient preserves the  (2,2) superconformal invariance.  Second, they offer a controlled setting for exploring aspects of string duality:  for example, one can ask about the interplay of CY mirror symmetry and such quotients along the lines recently discussed in~\cite{Cheng:2024noz}.  

These backgrounds are also of interest from a geometric perspective.  In the M-theory description as compactification on unorientable manifolds that nevertheless preserve supersymmetry, we are naturally led to consider a generalization and refinement of Berger's classification and its relation to manifolds with parallel spinors.\footnote{A review of these structures can be found in~\cite{Joyce:2000cm}.} Our main examples have holonomy $\SU(3)\times\Z_2 \subset \GO(7)$, but we also have somewhat more intricate cases with holonomy group $\SU(2)\rtimes\Z_2 \subset \GO(7)$.  A classification of the latter class of quotients may not be out of reach, and there is also motivation to study the case of $7$--dimensional compact flat manifolds with holonomy in $\GO(7)$ and parallel $\text{pin}^+$ pinors.

In addition these quotients highlight an intriguing geometric structure present in the moduli space of the original compactified theory, given by $\cM_V \times \cM_H$, with the vector moduli space a special K\"ahler manifold and the hyper moduli space a quaternion--K\"ahler manifold.   The quotient theory, a theory with $4$ supercharges naturally selects a subspace---the chiral multiplet moduli space--- $\cM_{\text{chi}} \subset \cM_V \times \cM_H$---a Hodge K\"ahler submanifold.  Since it is not particularly easy to find K\"ahler submanifolds in quaternion--K\"ahler spaces, it may be interesting to study these subspaces in detail.

The structure of this note is as follows: in section \ref{sec:typeII} we present the construction from the IIA perspective, and we turn to its M-theory lift in section \ref{sec:M}; we end with a discussion of other duality frames and variations on the idea in section \ref{sec:discussion}.  Two appendices contain some technical results.

\subsubsection*{Acknowledgements}
Some of this work was carried out while IVM was visiting LPTHE at Sorbonne University, as well as the Max Planck Institute for Gravitational Physics (AEI), and he is grateful to both institutions for hospitality and support.  RM would like to thank the Simons Center for Geometry and Physics for hospitality. We also thank P.~Cheng, M.~Dierigl, R.~Field, K.~Hori, S.~Theisen, and C.~Vafa for helpful discussions.

\section{The type II perspective}\label{sec:typeII}
Let $X$ be a smooth simply-connected CY $3$-fold  and denote by $\cC[X]$ the associated worldsheet theory in the RNS formalism.  This includes the internal SCFT for $X$, the degrees of freedom for the uncompactified $\R^{1,3}$ directions, as well as the superconformal ghosts.

  We suppose that $X$ admits a freely-acting K\"ahler isometry group $G$.  Any such action necessarily has finite order and must preserve the holomorphic form $\Omega$, so that the quotient $Y = X/G$ is another smooth CY manifold with a corresponding superconformal field theory  $\cC[Y]$.  The compactification of either IIA or IIB string on $X$ yields a four-dimensional effective theory with $8$ supercharges and a  well-known massless spectrum, with multiplicities determined by the  Hodge numbers $h^{1,1}(X)$ and $h^{1,2}(X)$.   In each case there is a universal sector, consisting of the $\cN=2$ supergravity multiplet and a hyper that contains the dilaton, the reduced NS-NS $2$-form, and R-R scalars (originating from $C_0$ and $C_2$ in IIB and from $C_3$ reduced on $H^{3,0}(X)$ in IIA).  The non-universal spectrum multiplicities are 
\begin{align}
				&& \text{vector}	&& \text{hyper} \nonumber\\
\text{IIA}		       	&&  h^{1,1}(X)	&& h^{1,2}(X)  \nonumber\\
\text{IIB}		 	&&  h^{1,2}(X)	&& h^{1,1}(X)
\end{align}
and the hypes are neutral with respect to the abelian vectors.  While there can be non-abelian gauge enhancement at special loci where some cycles become small, we will assume throughout this paper that we stay away from such loci.  

It is just as easy to describe the spectrum of the compactification on the quotient $Y$ at a generic point in the moduli space in terms of the orbifold theory $\cC[X]/G$.  Since $G$ is freely-acting, there are no marginal deformations in the twisted sectors, while in the untwisted sector we just keep the invariant deformations. The cohomology groups of $X$ decompose in representations of $G$:
\begin{align}
H^{p,q}(X) = H_{\text{triv}}(X)^{p,q} \oplus \oplus_{\rho \neq \text{triv}} H_{\rho}^{p,q}(X)~,
\end{align}
and $H^{p,q}(Y) \simeq H^{p,q}_{\text{triv}}(X)$.  

Type II string theory (either A or B) has a spacetime $\cF = \Z_2$ gauge symmetry that manifests itself in the RNS formalism as the global worldsheet symmetry $(-1)^{\bF_{\sleft}}$, which assigns charge $-1$ to the R-NS and R-R sectors while leaving the NS-NS and NS-R sectors invariant.\footnote{This is clear in string perturbation theory but is believed to be true non-perturbatively as well.  Perhaps the nicest argument for this is in the context of the IIA string, where the M-theory lift of this action corresponds to a parity action in M-theory:  a reflection of the dual circle, with corresponding action on the gravitino, as well as a change in the sign $C_3 \to -C_3$ of the $3$-form potential\cite{Vafa:1995gm,Diaconescu:2000wy}.}   This symmetry is closely related to the GSO projections that must be implemented when we construct the full world-sheet theory $\cC[X]$.\footnote{These have been recently discussed in~\cite{Delgado:2026qvy}, where it is pointed out that the required chiral GSO projections suffer from a global anomaly unless the target space $X$ is spin. }  There is also a right-moving version---$(-1)^{\bF_{\sright}}$, and the product of the two is the spacetime fermion number:  $(-1)^{\bF} = (-1)^{\bF_{\sleft}} (-1)^{\bF_{\sright}}$.

  We recall that the superstring chiral GSO projections involve the chiral fermion numbers of the internal SCFT which can be represented by $e^{\ii \pi J_{\sleft 0}}$ and $e^{\ii\pi J_{\sright 0}}$, where $J_{\sleft,\sright}$ are the currents associated to the $\GUL\times\GUR$ symmetry of the (2,2) SCFT, and the $J_{\sleft 0}$,$J_{\sright 0}$ denote the conserved charges.  Applying the construction to our $\cC[X]$ and using spectral flow to construct the Ramond sectors of the internal theory, we find that the Ramond sectors carry half-integral charges with respect to $J_{\sleft 0}$ and $J_{\sright 0}$, while NS-NS sector states of course carry integral charges.  It follows that we can represent $(-1)^{\bF_{\sleft,\sright}}$ by
\begin{align}
(-1)^{\bF_{\sleft} } & = e^{2\pi \ii J_{\sleft 0}}~,&
(-1)^{\bF_{\sright}} & = e^{2\pi \ii J_{\sright 0}}~.
\end{align}

\subsection{The orbifold symmetry and massless spectrum}

In light-cone gauge the physical massless states can be arranged into quartets organized according to sector and $\R^{1,3}$ spacetime helicity $h$, as well as the action of the $4$ spacetime supercharges that act non-trivially on the states.  We denote these by  $Q_{\sleft},Q_{\sright}$ and their conjugates $Q_{\sleft}^\ast,Q_{\sright}^\ast$; each pair is a doublet of the spacetime $\SU(2)$ R-symmetry, and while the former lower helicity, the latter raise it, in each case by $1/2$.  With that in mind, we find the following massless states, each arranged in a quartet
% according to 
%sectors in the pattern
\begin{equation}
\begin{tikzcd}[sep=small]
~ & h_{ \NSNS }& \\
%%%%%%%%
h_{\RNS} & ~ & h_{ \NSR} \\
%%%%%%%%
~ & h_{\RR }& 
\end{tikzcd}
\end{equation}
\begin{enumerate}
\item Supergravity multiplet.
\begin{equation}
\begin{tikzcd}[cramped,sep=small]
~ & +2 \ar[dl,"Q_{\sleft}"'] \ar[dr,"Q_{\sright}"] & \\
%%%%%%%%
+\ff{3}{2} \ar[dr,"Q_{\sright}"']  & ~ & +\ff{3}{2} \ar[dl,"Q_{\sleft}"] \\
%%%%%%%%
~ & +1 & 
\end{tikzcd}
\qquad
\begin{tikzcd}[sep=small]
~ & -2 \ar[dl,"Q^\ast_{\sleft}"']   \ar[dr,"Q^\ast_{\sright}"]  & \\
%%%%%%%%
-\ff{3}{2} \ar[dr,"Q^\ast_{\sright}"']   & ~ & -\ff{3}{2} \ar[dl,"Q^\ast_{\sleft}"]  \\
%%%%%%%%
~ & -1 & 
\end{tikzcd}
\end{equation}
We see that the content is organized according to $(-1)^{\bF_{\sleft}}$ in terms of $\cN=1$ multiplets with respect to the charges $Q_{\sright}$, $Q_{\sright}^\ast$:
\begin{align}
\left(\text{$\cN=2$ sugra}\right) \qquad = \qquad \left( \text{$\cN=1$ sugra}\right)_+  \oplus  \left( \text{$\cN=1$ gravitino}\right)_- ~.
\end{align}
\item Hypers.
\begin{equation}
\begin{tikzcd}[sep=small]
~ & 0  \ar[dr,"Q_{\sright}"] & \\
%%%%%%%%
+\ff{1}{2} \ar[dr,"Q_{\sright}"'] \ar[ur,"Q_{\sleft}"]  & ~ & -\ff{1}{2}  \\
%%%%%%%%
~ & 0  \ar[ur,"Q_{\sleft}"']& 
\end{tikzcd}
\qquad
\begin{tikzcd}[sep=small]
~ & 0  \ar[dr,"Q^\ast_{\sright}"] & \\
%%%%%%%%
-\ff{1}{2} \ar[dr,"Q^\ast_{\sright}"'] \ar[ur,"Q^\ast_{\sleft}"]  & ~ & +\ff{1}{2}  \\
%%%%%%%%
~ & 0  \ar[ur,"Q^\ast_{\sleft}"']& 
\end{tikzcd}
\end{equation}
Here we find that the content organizes according to
\begin{align}
\left(\text{$\cN=2$ hyper}\right) \qquad = \qquad \left( \text{$\cN=1$ chiral}\right)_+  \oplus  \left( \text{$\cN=1$ chiral}\right)_- ~.
\end{align}
\item Vectors.
\begin{equation}
\begin{tikzcd}[sep=small]
~ & 0  \ar[dr,"Q^\ast_{\sright}"] & \\
%%%%%%%%
+\ff{1}{2} \ar[dr,"Q^\ast_{\sright}"'] \ar[ur,"Q_{\sleft}"]  & ~ & +\ff{1}{2}  \\
%%%%%%%%
~ & +1  \ar[ur,"Q_{\sleft}"']& 
\end{tikzcd}
\qquad
\begin{tikzcd}[sep=small]
~ & 0  \ar[dr,"Q_{\sright}"] & \\
%%%%%%%%
-\ff{1}{2} \ar[dr,"Q_{\sright}"'] \ar[ur,"Q^\ast_{\sleft}"]  & ~ & -\ff{1}{2}  \\
%%%%%%%%
~ & -1  \ar[ur,"Q^\ast_{\sleft}"']& 
\end{tikzcd}
\end{equation}
These are organized as
\begin{align}
\left(\text{$\cN=2$ vector}\right) \qquad = \qquad \left( \text{$\cN=1$ chiral}\right)_+  \oplus  \left( \text{$\cN=1$ vector}\right)_- ~.
\end{align}
\end{enumerate}
We could take a quotient of the worldsheet theory by $(-1)^{\bF_{\sleft}}$---an operation that can already be done in ten dimensions and amounts to a flip of the GSO projection from the IIA one to the IIB one or vice versa, but that does not yield anything new:  we simply obtain a description of IIB string theory compactified on $X$, with additional gravitinos and matter returning in the twisted sector of the orbifold.  

To get something more interesting, we would like to combine the action of $(-1)^{\bF_{\sleft}}$ with the freely acting isometry $G$ so that the combined action is still freely acting.  Taking $G = \Z_3$ shows that this is not always possible:  the best we could get is the larger $\Z_3 \times \Z_2$ group, where the second factor is generated by $(-1)^{\bF_{\sleft}}$ by itself.  To avoid having a non-free action by $(-1)^{\bF_{\sleft}}$ we need to be able to assign a parity $\phi(g) \in\{ 0,1\}$ to each $g \in G$, so that we can obtain a new action by $\Gh \simeq G$ with 
\begin{align}
\Gh = \{ ~g (-1)^{\bF_{\sleft} \phi (g)}~~|~~ g \in G~\}~.
\end{align}
Equivalently, $G$ belongs to a short exact sequence
\begin{equation}
\label{eq:SymSeq}
\begin{tikzcd}
1 \arrow[r] & H  \arrow[r] & G \arrow[r,"\phi"] & \Z_2 \arrow[r] &1~.
\end{tikzcd}
\end{equation}
Thinking of $\Z_2$ as the additive $\Z/2\Z$ group, the subgroup $H \subset G$ consists of all elements $g$ with $\phi(g) = 0$, and it is not hard to see that it is both normal and maximal in $G$, and it contains the commutator subgroup: $\CO{G}{G}\subseteq H \subset G$. 
In what follows we will assume that the homomorphism $\phi$ is non-trivial.  Together $G$ and $\phi$ characterize the possible quotients and therefore the resulting four-dimensional theory, and we will see below that they similarly characterize unorientable $7$--dimensional M-theory backgrounds obtained as freely-acting quotients of $X \times S^1$.

Given such a group, we then wish to consider the orbifold worldsheet theory $\cC[X]/\Gh$.  We emphasize that this quotient is quite different from $\cC[X]/G$, where the action of the orbifold is purely in the internal sector.  This is not the case for $\cC[X]/\Gh$ since the Ramond sectors tie up all of the worldsheet degrees of freedom.

This is a subtle orbifold action.  For starters, it is defined on the full worldsheet theory with all the complications of ghosts, BRST--invariance, and non-compactness.   In addition, the orbifold action is chiral, so it possible that it may suffer from an anomaly.  We will give a spacetime argument below suggesting that $\Gh$ is anomaly--free, but it would be enlightening to either prove the statement directly in the worldsheet theory, or, if the claim is false, to characterize the anomaly--free subgroups of $\Gh$. 

Assuming that $\Gh$ is anomaly free, we can easily work out the resulting massless spectrum of the theory because the only contributions arise from the $\Gh$--invariant states in the untwisted sector.  To do so it helps to carry out an intermediate geometric step:  we first quotient $X$ by the maximal normal subgroup $H$ identified above to produce a Calabi-Yau manifold $Y_H = X/H$  with cohomology groups $H^{p,q}(Y_H) = H^{p,q}_{\text{triv,H}}(X)$.  Now $Y_H$ has the freely-acting isometry group $\Z_2 = \la r \ra $ with generator $r$ corresponding to the non-trivial element in $G/H$, and we decompose the cohomology groups of $X'$ into even and odd representations:
\begin{align}
H^{p,q}(Y_H) = H^{p,q}_{+}(Y_H) \oplus H^{p,q}_{-}(Y_H)~.
\end{align}
The quotient $Y_H/\la r\ra$ is exactly $Y = X/G$, with $H^{p,q}(Y) = H^{p,q}_{+}(Y_H)$.  

While the cohomology of $Y$ only knows about the $G$--invariant classes, the odd classes in $H^{p,q}_{-}(Y_H)$ do contribute to the massless spectrum, and looking back to our description of the multiplet structure we obtain the following for the invariant spectrum organized in $\cN = 1$ $d=4$ massless multiplets 
\begin{align}
\text{theory}  			&&  \text{universal sector}		&&  \text{vector multiplicity}	&&\text{chiral multiplicity} \nonumber\\
\text{IIA on}~  \cC[X]/\Gh	&&  \text{sugra}+\text{chiral}	&&  h^{1,1}_-(Y_H)~~~~~~	&& h^{1,1}_+(Y_H)+h^{1,2} (Y_H) \nonumber\\
\text{IIB on}~ \cC[X]/\Gh	&& \text{sugra}+\text{chiral}	&& h^{1,2}_-(Y_H)~~~~~~	&& h^{1,1}(Y_H) + h^{1,2}_+(Y_H)
\end{align}

\subsection{Compactification with holonomy} \label{ss:holonomy}
There is a an elegant spacetime point of view on the preceding worldsheet orbifold construction.   Assuming that $\cF = \la (-1)^{\bF_{\sleft}}\ra $ is indeed a gauge symmetry of the type II string, we identify the orbifold $\cC[X]/\Gh$ with the type II compactification on the non-simply connected space $Y$ with holonomy for $\cF$.  Such holonomies are classified by $\phi \in \Hom(\pi_1(Y), \Z_2)$.  Since any such homomorphism factors through the abelianization of $\pi_1(Y)$, we can equivalently think of this as $\phi \in  \Hom(H_1(Y,\Z),\Z_2) = H^1(Y,\Z_2)$.  Since $\pi_1(Y)= G$, this is exactly our choice of $\phi$ in the exact sequence defining the appropriate symmetry $G$.

This spacetime perspective strongly suggests that the orbifold construction should indeed be consistent:  we are merely turning on (discrete) background fields for a gauge symmetry of the theory, and while this can---and as we see above does---modify the low energy spectrum of the theory, it should not lead to an inconsistency.

\subsection{Examples}
In this section we present two examples of the construction for IIA on $\cC[X]/\Gh$.  For each case we take
take $X$ from~\cite{Braun:2009qy}.  $X$ is a Calabi-Yau manifold with $h^{1,1}(X) = 8$ and $h^{1,2}(X) = 44$ realized as a hypersurface in the toric variety $\text{dP}_3 \times\text{dP}_3$, where each $\text{dP}_3$ is a toric surface---$\C\P^2$ blown up at $3$ non-collinear points. 

 The beautiful and detailed analysis presented in~\cite{Braun:2009qy} characterizes the loci in the complex structure moduli space where $X$ admits a free $G$-action for several choices of $G$, including $\Z_3 \rtimes \Z_4$ and $\Z_{12}$.\footnote{This work was extended in a number of ways, including~\cite{Braun:2010vc}.
An overview of various Calabi-Yau constructions involving free quotients can be found in~\cite{Candelas:2016fdy}, as well as the more recent study~\cite{Gray:2021kax}.}  Focusing on the $\Z_{12}$ case, the cohomology group $H^{1,1}(X)$ decomposes into irreducible representations, denoted by $R_a$, for each of which the generator $x \in \Z_{12}$ acts by $x \cdot R_a = \zeta_{12}^a R_a$, where $\zeta_{12} = e^{2\pi\ii/12}$.  This representation is given by~\cite{Braun:2009qy}
\begin{align}
\label{eq:H11Decomp}
H^{1,1}(X) = R_0 \oplus R_2 \oplus R_3 \oplus R_4 \oplus R_6 \oplus R_8\oplus R_9 \oplus R_{10}~.
\end{align}
We will use this result to obtain two examples of $\cN=1$ compactifications.

\subsubsection*{An orbifold by $\Gh = \Z_2$}
Let $G = \Z_2 \subset \Z_{12}$ be the subgroup generated by $x^6$.   This is the symmetry subgroup of $\Z_{12}$ that remains once we make a generic complexified K\"ahler deformation by a modulus in the $R_2$ representation.

Each $R_a$ of $\Z_{12}$ is an irreducible $\Z_{2}$ representation, which we denote by $V_{0}$ and $V_{1}$, with $R_a = V_{a \mod 2}$, so that
\begin{align}
H^{1,1}(X) & = V_0^{\oplus 6} \oplus V_1^{\oplus 2}~.
\end{align}
The quotient $Y = X /G$ has Euler number $\chi(Y) = \chi(X)/ |G| = -36$, and therefore 
\begin{align}
h^{1,2}(Y) = -\frac{\chi(X)}{2 |G|} + h^{1,1}(Y) = 24~,
\end{align}
from which it follows that
\begin{align}
H^{1,2}(X) & = V_0^{\oplus 24} \oplus V_{1}^{\oplus 20}~.
\end{align}
Thus, if we take $\Gh = \Z_2$ to be generated by $x^6 (-1)^{\bF_{\sleft}}$, so that $Y_H = X$, then IIA on $\cC[X]/\Gh$ yields an $\cN=1$ theory with $2$ vectors and $51=1+6+44$ chiral multiplets (we include the axio-dilaton multiplet in this count.)

\subsubsection*{An orbifold by $\Gh = \Z_{12}$}
Now let us take $\Gh$ to be generated by $x (-1)^{\bF_{\sleft}}$, so that~(\ref{eq:SymSeq}) is non-trivial, with $H = \Z_6$ generated by $x^2$.  Every representation $R_a$ of $\Z_{12}$ is an irreducible representation of $H$---we denote these by $U_{0}, \ldots, U_{5}$---according to $R_a = U_{a \mod 6}$.  Thus, using~(\ref{eq:H11Decomp}) we find
\begin{align}
H^{1,1}(X) & = U_{0}^{\oplus 2} \oplus U_2^{\oplus 2} \oplus U_3^{\oplus 2} \oplus U_4^{\oplus 2}~.
\end{align}
It follows that $h^{1,1}(Y_H) = 2$, and
\begin{align}
h^{1,2}(Y_H) = -\frac{\chi(X)}{2 |H|} + h^{1,1}(Y_H) = 8~.
\end{align}
Since the two invariant classes descended from the representation $R_0 \oplus R_6 \subset H^{1,1}(X)$, we also see that $h^{1,1}_+(Y_H) = 1$ and $h^{1,1}_-(Y_H) = 1$.  These considerations are sufficient to determine the $\cN=1$ massless matter spectrum, which will now consist of $h^{1,1}_-(Y_H) = 1$ vector multiplet and $1+h^{1,1}_+(Y_H) +  h^{1,2}(Y_H) = 1+1+8=10$ chiral multiplets.

\section{The view from M-theory}\label{sec:M}
In the previous section we described in some detail the class of freely-acting orbifolds that combine $(-1)^{\bF_{\sleft}}$ with a discrete isometry in a type II compactification.  Now we re-examine the lift of such a construction in IIA to M-theory.

The starting point is an M-theory compactification on $\Zb = X \times S^1$---the lift of the IIA compactification on $X$---with the product Ricci-flat metric and $C_3 =0$.  We denote the  $2\pi$--periodic coordinate on $S^1$ by $\theta$.   $\Zb$ has a freely acting isometry $K$ isomorphic to the group $\Gh$ described above, with elements $g \in G$, which acts on a point $(p,\theta) \in \Zb$ by
\begin{align}
\label{eq:Kaction}
\kappa_g (p,\theta) = (g(p), (-1)^{\phi(g)} \theta)~,
\end{align}
and this is a symmetry of M-theory provided it is accompanied by $C_3 \to (-1)^{\phi(g)} \kappa_g^\ast C_3$.

\subsection{Properties of the quotient geometry}
The quotient geometry $Z = \Zb/K$ is smooth but necessarily non-orientable, since it acts on the volume form as
\begin{align}
\kappa_g^\ast \left( \ff{\ii}{8} \Omega\wedge \Omegab \wedge \ed \theta\right) = -\ff{\ii}{8} \Omega\wedge \Omegab \wedge \ed \theta~.
\end{align}
Nevertheless, $Z$ has a $\text{pin}^+$ structure, and the $\text{pin}^+$ bundle has covariantly constant sections.  We argue in Appendix \ref{sec:holonomy} that $Z$ has holonomy group $\Hol(Z) = \SU(3) \times \Z_2 \subset \GO(7)$.

The obstruction to a $\text{pin}^+$ structure on $Z$ is measured by the second Stiefel-Whitney class $w_2(Z)$~\cite{Kirby:1990pin}, and this can be difficult to verify in practice.\footnote{$\R\P^2$ is a classic example, reviewed in~\cite{Kirby:1990pin}, of a $\text{pin}^-$ manifold that cannot be given a $\text{pin}^+$ structure.}  However, we have a remarkably simple geometry at hand, and we will now argue that the $\text{pin}^+$ structure on the covering space descends to the quotient by presenting a $K$--equivariant description of the $\text{pin}^+$ bundle over the quotient geometry. 

We find it convenient to describe the quotient of the full $11$-dimensional spacetime $\Mb_{11} = \R^{1,3} \times X \times S^1$.  Since $\Mb_{11}$ is spin, it in particular admits a $\text{pin}^+$ structure, and we denote the corresponding vector bundle $\overline{\cP}_{+} \to \Mb_{11}$, with fibers $32$--component Majorana spinors.  We denote by $\Gamma^I$, $I=0,\ldots,10$ the corresponding set of Dirac matrices, with $\Gamma^{10}$ associated to the circle direction.

Given the action of the group $K$ on $\Mb_{11}$ we can construct a lift of this action to $\overline{\cP}_+ \to \Mb_{11}$, thereby giving it a $K$--equivariant structure.  That is, for every $g\in G$ we have a commutative diagram
\begin{equation}
\begin{tikzcd}
\kappa_g^\ast(\overline{\cP}_+) \ar[d] \ar[rr,"\hat{\kappa}_g"]& ~ & \overline{\cP}_+  \ar[d] \\
\Mb_{11} \ar[rr, "\kappa_g"] & ~& \Mb_{11}
\end{tikzcd}
\end{equation} 
and the lifts $\hat{\kappa}_g$, which act on the sections of the corresponding bundles, are compatible with the group action in the sense that for all $g_1$ and $g_2$ for all sections we have
\begin{align}
\hat{\kappa}_{g_1 g_2} = \hat\kappa_{g_2}  \kappa_{g_2}^\ast  \hat{\kappa}_{g_1}~.
\end{align}
Let $\sigma_g$ denote the isometry action of $g$ on $X$ and let $\sigmah_g$ denote a lift of this action to $\overline{\cP}_+$.  We know that such a lift exists since the action of $g$ preserves the spin structure on $X$, and of course it satisfies $\sigmah_{g_1g_2} =\sigmah_{g_2} \sigma^\ast_{g_2} \sigmah_{g_1}$.  Moreover, it also commutes with $\Gamma^{10}$.  We can set\footnote{We might think that there is a freedom to modify the lift by introducing an additional phase factor.  However, it is easy to check that this phase must be $\pm 1$ if the action is to be compatible with the Majorana condition that we impose on $\overline{\cP}_+$, and the results are not modified by this remaining choice of sign. }
\begin{align}
\label{eq:KappaLift}
\kappah_g = \sigmah_g \left(\Gamma^{10}\right)^{\phi(g)}~.
\end{align}
This is a well-defined map that is compatible with projections and the group structure, and therefore gives a lift of the action of $K$  to the total space $\overline{\cP}_+ \to \Mb_{11}$ that is compatible with the projection.  Since it includes elements with an odd number of $\Gamma$ matrices it is not compatible with a spin structure, but since $\kappah_g^2 = \text{id}$ we see explicitly that the quotient $\cP_+ = \overline{\cP}_+/K$ is a $\text{pin}^+$ bundle over $M_{11} = \Mb_{11}/K$.

It is also easy to see that the quotient preserves half of the supersymmetry of the original background.  To verify this we first parameterize the most general covariantly constant Majorana spinor on $\Mb_{11}$ in terms of a decomposition along the non-compact and compact directions.  Let $\gamma^\mu$ be a basis of Dirac matrices for $\R^{1,3}$, with chirality matrix $\gamma_5$, and let $\rho_1,\ldots,\rho_6$ be a basis of Dirac matrices for $\Spin(6)$, with chirality matrix $\rho_7$.  We can then take
\begin{align}
\Gamma^\mu & = \gamma^\mu \otimes\iden_8~,~~~ \mu = 0,\ldots, 3~,&
\Gamma^{3+a} & = \gamma_5 \otimes \rho_a~,~~~ a= 1,\ldots, 7~,
\end{align}
so that $\Gamma^{10} = \gamma_5 \otimes \rho_7$~.  There are charge conjugation matrices $\cC_4$ and $\cC_6$ such that $\cC_4 \gamma^\mu \cC_4^{-1} = -(\gamma^\mu)^{\text{t}}$ and $\cC_6 \rho_a \cC_6^{-1} =-(\rho_a)^{\text{t}}$, and $\cC_{11} = \cC_4 \otimes \cC_6$ then serves as the charge conjugation matrix for the $\Gamma^I$.  A Majorana spinor $\Psi$ in $11$ dimensions is defined by the condition $\Psi = \cC_{11} \Gamma^0 \Psi^\ast$, and the most general covariantly constant spinor on $\Mb_{11}$ is given by
\begin{align}
\Psi & = \eta \otimes \varepsilon_+ + (\cC_4 \gamma^0\eta^\ast) \otimes \cC_6 \varepsilon_+^\ast~,
\end{align}
where $\varepsilon_+$ is a normalized covariantly constant chiral spinor on $X$ pulled back to $\Mb_{11}$, and $\eta$ is an arbitrary constant Dirac spinor on $\R^{1,3}$.  The latter depends on $8$ arbitrary real parameters, and these correspond to the $8$ supercharges on $\R^{1,3}$ preserved by this background.

The preserved supersymmetries are then in 1:1 correspondence with the $\Psi$ that obey, for all $g \in G$,
\begin{align}
\Psi = \kappah_g \Psi = (\Gamma^{10})^{\phi(g)} \sigmah_g \Psi = (\gamma_5\otimes\rho_7)^{\phi(g)} \Psi~,
\end{align}
where in the last equality we used the form of $\Gamma^{10}$ and the assumption that the isometry action on $X$ preserves the chiral spinor $\varepsilon_+$.  Using the explicit form of $\Psi$ we then see that $\Psi = \kappah_g \Psi$ for all $g$ if and only if $ \eta = \gamma_5\eta$.  Thus, the degrees of freedom in $\eta$ are reduced to those of a chiral spinor on $\R^{1,3}$.

This discussion generalizes to the situation where $X$ admits $s>1$ covariantly constant chiral spinors, such as $X = T^6$ or $X = \text{K3} \times T^2$, in which case
\begin{align}
\Psi & = \sum_{a=1}^s \left( \eta_a \otimes \varepsilon_+^a + \text{c.c.} \right)~,
\end{align}
and
\begin{align}
\kappah_g \Psi & = \sum_{a=1}^s \left( (\gamma_5)^{\phi(g)} \eta_a \otimes \sigmah_g \varepsilon_+^a +\text{c.c.} \right)~.
\end{align}
As a simple case---we will meet an example shortly---we consider the situation where $\sigmah_g \varepsilon_+^a = (-1)^{\phi(g) p(a)} \varepsilon_+^a$, with $p(a) \in \{0,1\}$.  When this holds the quotient $M_{11}$ will preserve half of the supercharges of the original theory, with $\eta_a$ constrained to obey $\eta_a = (-1)^{p(a)} \gamma_5 \eta_a$.

\subsection{The $\SU(3)$ examples}
We can immediately see how this works for the two cases considered in the previous section, with $\Gh = \Z_2$ or $\Gh = \Z_{12}$, where the generators are
\begin{align}
\Gh = \Z_2 &:  
\kappa(p,\theta) = (x^6(p),-\theta)~, &
\Gh = \Z_{12} &:
\kappa(p,\theta) = (x(p),-\theta)~.
\end{align}
For each of these the quotient preserves half of the supersymmetries, as we already showed in the IIA description, but we can now also reproduce the massless spectrum from the M-theory point of view.  For simplicity we will just discuss the $\Gh = \Z_2$ case.   The bosonic massless fields arise from the metric moduli and from reduction of $C_3$ on harmonic forms on $\Zb$.\footnote{The Kaluza-Klein vector from the metric is projected out.}  From the former we obtain (we count real degrees of freedom)
\begin{align}
N_{\text{metric}} & = h^{1,1}_{+}(X) + 2 h^{1,2}_{+} (X)+1
\end{align}
scalars, with the last contribution coming from the radius of the M-theory circle.   To describe the reduction of $C_3$ we observe that
\begin{align}
H^1(\Zb,\R) & = (\R)_-~,&
H^2(\Zb,\R) & = (\R^{h^{1,1}_+(X)})_+ \oplus (\R^{h^{1,1}_-(X)})_-~,
\end{align}
as well as
\begin{align}
H^3(\Zb,\R) 	& = H^3(X,\R) \oplus H^2(X,\R)\otimes H^1(S^1,\R) 
\nonumber\\
			& = (\R^{2+2h^{1,2}_+(X)+h^{1,1}_-(X)})_+ \oplus (\R^{h^{1,1}_+(X)+2h^{1,2}_-(X)})_-
\end{align}
Since our action flips $C_3 \to -C_3$, we see that $C_3$ yields
\begin{align}
V = h^{1,1}_-(X)
\end{align}
gauge bosons, which matches the expected number of vector multiplets from our IIA computation, and
\begin{align}
N_{C_3} & = 1 + h^{1,1}_+(X) + 2 h^{1,2}_-(X)
\end{align}
scalars, where the first one arises by dualizing the spacetime $2$-form from the reduction of $C_3$ on $H^1(\Zb,\R)$.  Adding up the scalars, we find $N_{\text{chi}} = 1 + h^{1,1}_+(X) + h^{1,2}(X)$, as we already saw in the IIA description.

\subsection{Two $\cN=2$ quotients}
In this section we examine two closely related quotient geometries in M-theory.  Each quotient reduces the supersymmetry from $16$ to $8$ supercharges, and, remarkably, while the two have identical massless spectra, one is orientable, while the other is not.

We start with the geometry $\Zb = \Mb \times T^3$, where $\Mb$ is a K3 surface with Einstein metric tuned to the Enriques locus.  That is, the metric on $g$ is chosen so that $\Mb$ admits the Enriques involution $\sigma$---a freely-acting K\"ahler isometry which acts on the de Rham cohomology of $\Mb$ as
\begin{align}
H^{0}(\Mb) & = H^4(\Mb) = (\R)_+~, 
\nonumber\\
H^{2}(\Mb) & = (\R^{10})_+ \oplus (\R^{12})_-~,
\nonumber\\
H^{2}(\Mb) \otimes_{\R} \C & = 
\underbrace{H^{2,0}(\Mb)}_{\simeq(\C)_-} \oplus \underbrace{H^{1,1}(\Mb)_{+}}_{\simeq (\C^{10})_+} \oplus \underbrace{H^{1,1}(\Mb)_-}_{\simeq (\C^{10})_-} \oplus  \underbrace{H^{0,2}(\Mb)}_{\simeq(\C)_-}~.
\end{align}
This restricts the choice of metric to a $\dim_{\R} = 30$ sub-locus in the $\dim_{\R} = 58$ moduli space of Einstein metrics on $\Mb$.

While the quotient just by $\sigma$ does not produce a geometry suitable for M-theory ($\Mb/\la \sigma\ra$ is $\spin_{\text{c}}$), we can combine it with an action on $T^3$ to construct suitable quotient geometries.  Denote by $\rho_i$, $i=1,2,3$, the reflection in the $i$-th circle of $T^3$.  We consider three distinct $K= \Z_2$ quotients, with generators given by, respectively,
\begin{align}
\kappa_1 & = \sigma \rho_1~,&
\kappa_{2} & = \sigma \rho_1 \rho_2~,&
\kappa_{3} & = \sigma \rho_1 \rho_2 \rho_3~.
\end{align}
The first of these does not preserve supersymmetry:  $\sigmah$ acts on the spinor bundle with factors of $\pm \ii$, and these cannot be compensated by a single reflection.  The second leaves the third circle invariant, and the quotient geometry is $Y \times S^1$, where $Y$ is the Enriques Calabi-Yau $3$-fold, with $h^{1,1}(Y) = h^{1,2}(Y) = 11$ \cite{Ferrara:1995yx}.  We recall that the massless gauge bosons arise by reducing $C_3$ on $H^2(Y)$ (the Kaluza-Klein gauge boson is the graviphoton in the supergravity multiplet), while the $70$ real scalars arise as follows:
\begin{enumerate}
\item  $30$ scalars are associated to deformations of the K3 metric on the Enriques locus;
\item  $4$ scalars are associated to the torus:  $3$ to deformations of the reflected $T^2$,  $1$ to the radius of the unreflected $S^1$;
\item the remaining $36$ scalars arise by reducing $C_3$ on the cohomology of $Y \times S^1$, and this yields $11+24+1$ scalars, with the $11$ coming from reduction on $H^{2}(Y) \otimes H^1(S^1)$, the $24$ from reduction on $H^3(Y)$, and the last scalar is the dual of the spacetime $2$-form obtained by reducing $C_3$ on $H^1(S^1)$.
\end{enumerate}

Let us now consider the remaining quotient $\Zb / K$, where $K$ is generated by $\kappa_3$.  $Z = \Zb/K$ is unorientable, but by the preceding arguments we know that the quotient will still preserve $8$ supercharges.  It is straightforward to determine the massless spectrum: we just decompose the cohomology of $\Zb$ into representations of $K$ and recall that the symmetry action acts on $C_3$ with an extra sign:  $C_3 \to -\kappa^\ast_3 C_3$. The $\Zb$ de Rham cohomology decomposes as
\begin{align}
H^0(\Zb) & = (\R)_+~, \nonumber\\
H^1(\Zb) & = H^1(T^3) = (\R^3)_-~, \nonumber\\
H^2(\Zb) & = H^2(\Mb)\oplus H^2(T^3)  = (\R^{10})_+ \oplus (\R^{12})_- \oplus (\R^{6})_+~, \nonumber\\
H^3(\Zb) & = H^2(\Mb) \otimes H^1(T^3) \oplus H^3(T^3) = (\R^{30})_- \oplus (\R^{36})_+ \oplus (\R)_{-}~,
\nonumber\\
\vdots & \nonumber\\
H^7(\Zb) & = H^4(\Mb) \otimes H^3(T^3) = (\R)_-~.
\end{align}
We therefore obtain $12$ gauge bosons by reducing $C_3$ on $(\R^{12})_- \subset H^2(\Zb)$:  one of these is the graviphoton, and the rest reside in $11$ vector multiplets.   Note that the Kaluza-Klein gauge bosons from the metric reduced on $T^3$ are odd and do not contribute to the massless spectrum.  The scalars arise as follows:
\begin{enumerate}
\item $30$ again describe the deformations of K3 metric along the Enriques locus;
\item $6$ correspond to deformations of the flat metric on $T^3$;
\item $31$ come from reducing $C_3$ on $H^3(\Zb)_-$, and the remaining $3$ are dual to the spacetime $2$-forms obtained by reducing $C_3$ on $H^1(\Zb)_-$.
\end{enumerate}
So, the somewhat surprising conclusion is that the massless spectrum for M-theory compactified on the unorientable manifold $Z = \Zb/K$ is a theory with $8$ supercharges, $11$ vectors, and $12$ hypers---exactly the massless content of M-theory compactified on the oriented manifold $Y\times S^1$, where $Y$ is the Enriques Calabi-Yau manifold.  

We do not know whether the two theories are equivalent or are instead different UV completions of the same theory, providing a lower-dimensional example of global variants familiar from $9$ and $8$ dimensions~\cite{Hellerman:2005ja,Aharony:2007du,Montero:2022vva}.  Geometrically, they are distinguished by the holonomy group:  the holonomy of $Y \times S^1$ is $\SU(2) \rtimes \Z_2 \subset \SO(7)$, while that of $Z$ is not contained in $\SO(7)$.  Instead, $\text{Hol}(Z) = \SU(2) \rtimes \Z_2 \subset \GO(7)$.  We discuss these holonomy groups in more detail in Appendix \ref{sec:holonomy}.

\subsection{An oriented example with unoriented consequences} \label{ss:compactflat}
Our examples so far have all had a continuous component to the holonomy group.  However, even if the holonomy group is discrete, and the geometry is just a compact flat Riemannian manifold, it is interesting to interpret the presence of $(-1)^{\bF_{\sleft}}$ in the quotient geometrically.  To illustrate this, we consider M-theory compactified on $Z= (T^4\times S^1)/\Z_2$, where the $\Z_2$ action is generated by
\begin{align}
\kappa(\theta_1,\theta_2,\theta_3,\theta_4,\theta_5) = (-\theta_1,-\theta_2,-\theta_3,-\theta_4,\theta_5 + \pi)~.
\end{align}
We can now apply the lesson from~\cite{Vafa:1995gm} on realizing these symmetry actions in the reduction to IIA:  a reflection of the M-theory circle means the orbifold action has a factor of $(-1)^{\bF_{\sleft}}$, while each additional reflection in M-theory of a $\theta_i$ reduces as $\cR_i \Pi$, where $\cR_i$ represents the reflection in IIA, while $\Pi$ is worldsheet parity.   In our case there are two simple possibilities for M-theory reduction:  we can reduce on the ``shifted circle'' $\theta_5$ or on one of the reflected circles, say $\theta_4$.   

Reduction on $\theta_4$ can be interpreted as a $\Z_2$ IIA orbifold of $T^4$ with generator
\begin{align}
\label{eq:flatexamplegeneratorIIA}
\gh & = \cR_1 \Pi \cR_2 \Pi \cR_3 \Pi (-1)^{\bF_{\sleft}} \cS_5~,
\end{align}
where $\cS_5$ is an order $2$ shift orbifold action on the $5$-th circle.
This is an orientifold action but without orientifold planes, of the sort recently discussed in~\cite{Cheng:2023owv}.
The compactification manifold is a smooth unorientable manifold with a $\text{pin}^-$ structure, but the insertion of $\Pi (-1)^{\bF_{\sleft}}$ renders the orientifold supersymmetric and well-defined.

On the other hand, reduction on the shifted circle direction leads to a singular oriented compactification manifold (namely $T^4/\Z_2$), but with a non-trivial flat bundle for the Ramond-Ramond potential $C_1$.  By applying a single T-duality followed by an S-duality in the IIB description this can be related to a flat $B$-field gerbe background discussed in some detail in~\cite{Cheng:2022nso}.

This example is just a glimpse of the vast class of compact flat Riemannian manifolds, and it would be interesting to extend the classification results such as those obtained in~\cite{Lutowski_2015,Dekimpe:2009kfm} to unoriented manifolds with $\text{pin}^+$ structure, and to explore the implications of string dualities for the resulting class of string/M-theory vacua.

\section{Discussion}\label{sec:discussion}

Having described a class of---we hope---intriguing string vacua, it is natural to consider how this class can be interpreted in other duality frames.   The aim of this section is to point out some first steps in developing this kind of understanding.

\subsection{Relations between IIA and IIB quotients}
As we mentioned at the start of our discussion combining $(-1)^{\bF_{\sleft}}$ with a free action $G$ on a Calabi-Yau $3$-fold $X$ can be done in either the IIA or the IIB description.  The resulting theories typically have no simple relation.  Since mirror symmetry relates the IIA string compactified on $X$ to the IIB string compactified on the mirror manifold $X^\circ$, we do have a way of constructing the mirror description of IIA on $\cC[X]/\Gh$ because by construction $\cC[X^\circ]$ admits a mirror global symmetry:  $G^{\circ} = \mu G \mu^{-1}$, where $\mu : \cC[X] \to \cC[X^\circ]$ is the mirror isomorphism.  We can then combine $G^{\circ}$ with $(-1)^{\bF_{\sleft}}$ into an action $\Gh^\circ$ and attempt to construct the IIB orbifold $\cC[X^\circ]/\Gh^\circ$.  But, there is a complication:  the action of $G^\circ$ on $X^\circ$ will not be free.  Since we know the expected answer from the IIA description, understanding the details of the $\cC[X^\circ]/\Gh^\circ$ orbifold directly should be possible, and it may provide an excellent point of departure for generalizing the construction to non-free actions.

In the case of compact flat backgrounds T-duality can be used to relate the IIA and IIB descriptions, but even in this seemingly simple setting following the duality map can be instructive.  Returning to our IIA example from section~\ref{ss:compactflat}, we obtain a dual action for the IIB orbifold of $T^4$ by applying T-duality along the third circle.\footnote{We collect some properties of T-duality, worldsheet parity and related matters in Appendix \ref{sec:T}.}  With $\cT_i$ denoting the T-duality along the $i$-th circle, which we take to be a reflection of the right-moving movers with a consistent lift to the Ramond sectors, we find that the IIB dual of the action~(\ref{eq:flatexamplegeneratorIIA}) is
\begin{align}
\gh^{\circ} = \cT_3 \gh \cT_3^{-1} & = \cR_{1} \cR_{2} \Pi (-1)^{\bF_{\sleft}} \cS_5 ~,
\end{align}
The geometry is therefore $T^3/\Z_2 \times S^1$, where the $\Z_2$ action on $T^3$ is generated by
\begin{align}
\kappa(\theta_1,\theta_2,\theta_5) = (-\theta_1,-\theta_2,\theta_5+\pi)~.
\end{align}
Without the accompanying worldsheet factor of $\Pi(-1)^{\bF_{\sleft}}$ this would not be supersymmetric, but of course this orientifold preserves $16$ supercharges.   This has a spacetime interpretation.  The fermions of the IIB string transform non-trivially under $\SL(2,\Z)$, and $\Pi(-1)^{\bF_{\sleft}}$ generates a perturbative subgroup $\Z_4 \subset \text{Pin}^+\GL(2,\Z)$---the lift of $\SL(2,\Z)$ and IIB perturbative symmetries to the full set of massless fields in IIB supergravity~\cite{Debray:2021vob}.  Similarly, $\cR_1 \cR_2$ generates a $\Z_4$ subgroup of $\Spin(3)$.  Their combination $\gh^{\circ}$ yields a $\Z_2$ action on the spacetime fermions, and the orientifold has the spacetime interpretation as compactification on the geometry with holonomy for $\gh^{\circ}$ along the appropriate homotopy class $[\gamma] \in \Z_2 \subset \pi_1( T^3/\Z_2 \times S^1)$.

\subsection{Geometrization of IIB orbifolds and F-theory}
Given the M-theoretic geometrization of the $g(-1)^{\bF_{\sleft}}$ action on $X$ in IIA, it is natural to ask about a similar geometrization in the IIB context via an F-theory lift.  Perhaps the simplest lifts of this sort involve the combination of a free geometric action on the compactification manifold and the worldsheet symmetry $(-1)^{\bF_\sleft} \Pi$.  Since the latter is identified with the center of the $\SL(2,\Z)$ duality group of IIB, such orbifolds have a ``flat F-theory'' lift to a quotient $(X \times T^2)/ K$, where $K$ acts on the $T^2$ fiber by a reflection of both cycles~\cite{Cheng:2023owv}.\footnote{For instance, this yields the F-theory lift for IIB compactification on the Enriques surface, where the base of the elliptic fibration is K\"ahler but the canonical bundle is non-trivial---its first Chern class is a $\Z_2$--torsion class.  Thus, the base manifold (Enriques surface) is not spin, and the action of  $\Pi (-1)^{\bF_{\sleft}}$ is needed to preserve spacetime supersymmetry. Note that there is no direct $4$-fold analogue to Enriques $3$-fold, as discussed in~\cite{Cheng:2023owv}.}  The duality group of IIB is in fact larger, and $\Pi$ and $(-1)^{\bF_{\sleft}}$ act as two independent reflections of the F-theory $T^2$~\cite{Tachikawa:2018njr}, so that these elements provide the lift we seek: in F-theory the quotient is again of the form $(X \times T^2)/ K$, but now $K$ acts on the $T^2$ fiber by a single reflection, leading to F-theory compactification on an unorientable manifold.  In the examples discussed above this can be realized by taking the F-theory limit of the M-theory compactification on this eight-dimensional (product) pin$^+$ manifold.

\subsection{Other smooth quotients and their consistency}
Let us finally consider a few related constructions that combine the free action by $G$ on a Calabi-Yau manifold $X$ with other stringy symmetries.

\subsubsection*{Barely $G_2$ manifolds and quotients with $\text{pin}^-$ structure}  
Recall that a ``barely $\GG_2$'' manifold is a seven-dimensional space $(X\times S^1)/K$, where $X$ is a smooth Calabi-Yau $3$ fold that admits an antiholomorphic involution $\sigma$ that acts on the K\"ahler and holomorphic forms by $\sigma^\ast J = -J$ and $\sigma^\ast \Omega = \Omegab$.   When $X$ is equipped with a product Ricci-flat metric, $X\times S^1$ admits a family of $\GG_2$ structures, with associative form labeled by a real parameter $\alpha$:
\begin{align}
\Phi_\alpha & = \ed \theta   \wedge  J + \Rea \left( e^{\ii \alpha} \Omega\right)~.
\end{align}
Setting $Z = (X\times S^1)/K$, where $K$ is generated by $\kappa (p,\theta) = (\sigma(p),-\theta)$, we see that $K$ preserves a $\GG_2$ structure with associative form $\Phi_0$.   Note that while the anti-holomorphic involution $\sigma$ on a $3$-fold $X$ is orientation-reversing, the combined action on $X\times S^1$ preserves orientation.  If $\sigma$ acts freely on $X$, $Z$ is a manifold of holonomy $\SU(3) \rtimes \Z_2$,\footnote{Otherwise the orbifold is a starting point for constructing manifolds of full $\GG_2$ holonomy by a resolution of singularities compatible with the $\GG_2$ structure.  A review of the methods and challenges associated to such resolutions is given in~\cite{Joyce:2017nc}.}---a barely $\GG_2$ manifold.    We can compactify M-theory on such a $Z$ and of course interpret the result in IIA compactified on $X$ as an orbifold that combines the anti-holomorphic involution with $(-1)^{\bF_{\sleft}}$.   

This construction is analogous to the well-understood compactification of M-theory on the Klein bottle, where we replace the shift circle with a Calabi-Yau $3$ fold $X$ acted upon by $\sigma$.  It is tempting to push the analogy further:  might we not be able to instead replace the reflected circle by $X$?  If $\sigma$ acts freely on $X$ then we need not even accompany it by a shift in $S^1$, and it appears that we obtain an M-theory compactification on the unorientable manifold $X/\la \sigma\ra \times S^1$.   This is suspicious, since it suggests that the smooth manifold $X/\la\sigma\ra$ leads to a well-defined background for IIA, in contradiction with worldsheet analysis~\cite{Delgado:2026qvy}.    If $\sigma$ does not act freely, we can accompany it by a half--period shift in the M-theory circle, leading to a free action on the $7$--manifold for any anti-holomorphic involution $\sigma$.  In the IIA description this is interpreted as the introduction of a flat background for the RR $1$-form potential, making a worldsheet consistency analysis more challenging, but the M-theory perspective suggests that all of these specious backgrounds are inconsistent.

The flaw in the M-theory construction is easy to spot:  because locally the action of $\sigma$ involves a reflection of $3$ coordinates, the analogue of the lift of the action to the Majorana spinor bundle given in~(\ref{eq:KappaLift}) involves a product of $3$ $\Gamma$ matrices, say $\Gamma^5\Gamma^7\Gamma^9$, and the resulting action squares to $-1$.  This means we obtain a $\Z_2$--equivariant $\text{pin}^-$ structure on $\overline{\cP}_+ \to M_{11}$, but this is not consistent in M-theory.  We stress that the issue is not merely preservation of supersymmetry but rather consistency:  since the M-theory background manifold is smooth and has a flux-free large radius limit, the supergravity limit should give a reliable description of the consistency conditions, and these require a $\text{pin}^+$ structure.\footnote{The situation is more subtle at finite volume.  For instance, T-duality implies that M-theory can be compactified on the $\text{spin}_{\text{c}}$ manifold $\C\P^2\times T^2$~\cite{Duff:1998us}, albeit with flux, as well as membrane winding states contributing to the spectrum and being responsible for the supersymmetry of the background.}  The same inconsistency also afflicts the more general construction where the anti-holomorphic involution is accompanied by a shift in the circle.

\subsubsection*{Consistent non-supersymmetric quotients}
Freely-acting supersymmetry breaking quotients of $T^n$, where a $(-1)^{\bF}$ action is accompanied by shifts, were recently discussed in~\cite{Baykara:2026jzs} as possible backgrounds of M-theory.  As suggested there, we may consider replacing the torus by a Calabi-Yau $3$-fold $X$ that admits a free action by $G$ while also replacing $(-1)^{\bF_{\sleft}}$ in our discussion above by $(-1)^{\bF}$.\footnote{Recent work has also given an M-theory interpretation to the 0A string --- an orbifold of IIA by $(-1)^{\bF}$---as M-theory on the non-geometric space $(S^1 \vee S^1)$~\cite{Baykara:2026gem,Altavista:2026evd,Baykara:2026vdc}.  The free quotients we consider are still backgrounds of the IIA string:  in the decompactification limit we recover the supersymmetric theory.}

If we just consider the $(-1)^{\bF}$ action in an M-theory compactification as a truncation to the bosonic fields, we obtain an obvious contradiction, since a brutal removal of the spacetime fermions ruins the delicate conspiracy between the Rarita-Schwinger determinant and the Chern-Simons couplings that renders the partition function well--defined ~\cite{Witten:1996md, Freed:2019sco}.  On the other hand, taking $M_{11} = M_5 \times X$ as a background for M-theory truncated to bosonic fields would naively yield a five-dimensional theory that appears to be consistent (in the sense of having a well-defined partition function), since only an integrality of a certain cubic form in five dimensions would be required \cite{Cheng:2025ikd}.  In a little more detail,  for a five-dimensional theory with $n$ vector fields $A^i$ with field strength $F^i = \ed A^i$ ($i=1,...,n$) and Chern-Simons couplings 
\begin{equation}
\int_{M_5}\left\{  -\frac16 c_{ijk} A^i \wedge F^j \wedge F^k - \frac{a_i}{48} A^i \wedge p_1(M_5) \right\},
\end{equation}
this integrality condition, and hence the existence of a well-defined partition function, is valid provided that all components of $c_{ijk}$ and $a_i$ are integers and 
\begin{equation}
\label{eq:intcube}
\frac16 c_{iii} + \frac{a_i}{12} \in \Z \quad \mbox{and}  \quad \frac12(c_{iij} + c_{ijj}) \in \Z
\end{equation}
for any value of $i$ and $j$. While for minimal supergravity in five dimensions coupled to $n-1$ vector multiplets it is equivalent to BPS $(0,4)$ strings  having central charge $c_\sright$ divisible by six, the condition does not rely on supersymmetry and is necessary and sufficient for a well-defined one-loop partition function.

On the other hand, combining the $(-1)^{\bF}$ action with a freely-acting isometry of $X$ entirely changes the perspective, since we can now think of the construction as a compactification with a holonomy for the $(-1)^{\bF}$ gauge symmetry of the $11$-dimensional theory supported by non-trivial cycles in the quotient space along the lines discussed in section~\ref{ss:holonomy}.   At the level of the spectrum of the $5$--dimensional theory this gives masses to a subset of the fields but has no effect on the integrality conditions  \eqref{eq:intcube} that must be obeyed by the Chern-Simons terms for a consistent theory.   Thus, from either the $11$--dimensional or $5$--dimensional perspective it appears that the result is a consistent theory.   Of course determining whether or not it is indeed an M-theory vacuum is much more challenging.  It would be extremely interesting to generalize the duality group--based arguments of~\cite{Baykara:2026jzs} to this situation.

\bigskip
\noindent
To conclude, compactifications on unorientable supersymmetric seven-manifolds and the associated  ${\mathcal N}=1$ four-dimensional theories deserve further study. While interesting on their own, these also point to an existence of an unexplored refined structure of the moduli space of ${\mathcal N}=2$ theories obtained form compactifications on Calabi-Yau manifolds admitting a freely acting K\"ahler involution.  Besides exploring these structures in more detail, a more ambitious goal is to generalize the analysis to genuine orbifolds with fixed points, but even finding necessary and sufficient conditions for consistency of such theories will require new tools.   Another important open problem is to characterize---and possibly to compute---the non-perturbative corrections to the ${\mathcal N}=1$ superpotential.  Given the CY geometry underlying the construction, these corrections may be easier to characterize than such effects in generic M-theory compactification on $\GG_2$ manifolds.

\appendix

\section{Holonomy groups and orientation}\label{sec:holonomy}

\subsection{A geometric review}
We begin with a brief review of Riemannian holonomy, following~\cite{Besse:1987pua,Joyce:2000cm}.   Let $(X,g)$ be a compact Riemannian manifold of dimension $n$ equipped with a metric $g$.  By using the Levi-Civita connection to parallel transport vectors on closed loops $\gamma : [0,1] \to X$ with $\gamma(0) =\gamma(1) = x \in X$ we generate the holonomy group $\Hol_x(X,g) \subset \GL(n,\R)$ with elements
\begin{align}
\Hol_x(X,g) = \{ P_{\gamma} ~~|~~ \gamma \in \text{all loops based at } x\}~,
\end{align}
where $P_\gamma \in \GL(n,\R)$ is the isomorphism $P_\gamma : T_x X \to T_x X$ obtained by parallel transport,
and the group product is obtained by composing the loops.
The restricted holonomy group is a subgroup of $\Hol_x(X,g)$ generated by the homotopically trivial loops:
\begin{align}
\Hol_x^0(X,g) = \{ P_{\gamma}~~|~~\gamma \in \text{all loops based at } x ~~\text{with}~~ [\gamma]  = \text{id} \in \pi_1(X)\}~.
\end{align}
The dependence on the base point $x$ can be dropped because a change of base point merely conjugates the group inside $\GL(n,\R)$.  So, we characterize the subgroup up to conjugation and drop the $x$ label.  In what follows we will also leave the Riemannian metric $g$ implicit in the notation.  In all of our examples it will be a Calabi-Yau metric, possibly including a flat component when the manifold is reducible.  Since the connection preserves the metric we in fact have $\Hol(X) \subseteq \GO(n)$.

These groups fit into the following diagram with short exact row and short exact column:
\begin{equation}
\label{eq:HolonomyDiagram}
\begin{tikzcd}
~			&		~		&	~			&  1 \ar[d]& ~\\
~			&		~		&	~			&  N \ar[d]	& ~\\
~			&		~		&	~			&  \pi_1(X) \ar[d,"\varphi"]	\ar[dl,"f"']& ~\\
1 \ar[r]		&\Hol^0(X) \ar[r,"i"] 	& \Hol(X) \ar[r,"p"]	&Q \ar[r] \ar[d]	&1 \\
~			&		~		&	~			& 1			&~
\end{tikzcd}
\end{equation}
In other words, $\Hol^0(X) \subset \Hol(X)$ is a normal subgroup, and there is a surjective homomorphism $\varphi: \pi_1(X) \to \Hol(X)/\Hol^0(X)\simeq Q$.
The map $f$ is a lift of $\varphi$ to $\Hol(X)$, where for each class $[\gamma] \in \pi_1(X)$ we pick a representative loop $\gamma$, and then set $f([\gamma]) = P_{\gamma}$.

We make a few remarks that are probably quite unnecessary for geometers but may be useful to physicists (as least we found them to be so when thinking about these matters).
\begin{enumerate}
\item  $T^n$ has a non-trivial fundamental group $\pi_1(T^n) = \Z^n$, but $\Hol(T^n) = \Hol^0(T^n) = 1$.  So $N$ can certainly be non-trivial.
\item  More generally, the holonomy of a Riemannian compact flat manifold $X =\R^n/\pi_1(X)$ has the following structure~\cite{Charlap:1986fcm}:
\begin{equation}
\begin{tikzcd}
1 \ar[r]	& N \ar[r]	& \pi_1(X)\ar[r] & \Hol(X) \ar[r] &1
\end{tikzcd}~.
\end{equation}
Here $N$ is a free abelian normal subgroup generated by $n$ linearly independent translations.
That fits nicely with the general story, since $\Hol^0(X) = 1$ in this case.
\item $\Hol(X)$ is always contained in the normalizer subgroup of $\Hol^0(X)$, realized as a subgroup of $\GO(n)$, and if $X$ is oriented, then the subgroup is contained in $\SO(n)$. 
\item The Berger classification of holonomy groups of irreducible oriented Riemannian manifolds is a statement about the restricted holonomy group, or, equivalently, the holonomy group of the universal cover of $X$, $\Xb$.
\end{enumerate}

\subsection{The holonomy of an Enriques surface}
Consider the Enriques surface $M = \Mb/\la \sigma\ra$, where $\sigma$ is the Enriques involution discussed above in the text.  Since $M$ is a Ricci-flat K\"ahler manifold it must be that
\begin{align}
\Hol^0(M) = \SU(2) \subset \Hol(M) \subsetneq \GU(2)~.
\end{align}
We also have $\pi_1(M) = \Z_2$, and thus the only consistent possibility for~(\ref{eq:HolonomyDiagram}) is that $N=1$, so that
\begin{equation}
\begin{tikzcd}
1 \ar[r]	& \Hol^0(M) \ar[r]	& \Hol(M)\ar[r] & \pi_1(M) \ar[r] &1
\end{tikzcd}~.
\end{equation}
Now let $\cT \in \Hol(M)$ be any element that maps to the generator of $\pi_1(M)$; we can represent $\cT$ in the fundamental representation of $\GU(2)$, and it must be that
\begin{align}
\cT & = e^{\ii \psi} \cS~,& \cT^2 & = e^{2\ii\psi} \cS^2 \in \SU(2)~,
\end{align}
where $\cS \in \SU(2)$.  Since $\cT \not\in \SU(2)$, we can choose $\cS$ so that $\cT = \ii \cS$, or, equivalently $\cT = \sigma_3 \cS'$, where $\cS' = \ii \sigma_3 \cS$, and $\sigma_3$ is the usual $2\times 2$ Pauli matrix.  This is enough to show that $\Hol(M) = \SU(2) \rtimes \Z_2$, where every element is uniquely represented as $\sigma_3^a \cS$, where $\cS \in \SU(2)$ and $a = 0,1$. Equivalently\footnote{This discussion closely follows the presentation explained by Robert Bryant in a post on MathOverflow from 2013.  Every element in $\Hol(M)$ can also be written as $A=\ii^{a} \cS$, with $a \in \Z/4\Z$ and $\cS \in \SU(2)$, with the identification $(a,\cS) \sim (a+2,-\cS)$, and that leads to the presentation of $\Hol(M)$ given in~\cite{Besse:1987pua} as $\SU(2) \rtimes_{\Z_2} \Z_4$.}
\begin{align}
\Hol(M) = \{ A \in \GU(2)~~|~~ \det A = \pm 1\}~.
\end{align}

\subsection{The holonomy of the Enriques Calabi-Yau $3$-fold}
We now consider $Y = (\Mb \times T^2) / \la \kappa_2\ra$, where $\kappa_2 = \sigma \rho_{1} \rho_{2}$ combines the Enriques involution on $\Mb$ with a reflection of both circles on $T^2$.  We have $\pi_1(Y) = \Z^2\rtimes \Z_2$.  To see this consider the universal cover $\Yb = \Mb \times \C$, so that $\pi_1(Y)$ can be represented by the generators $T_{m,n}$ and $R$ acting on points $(x,z) \in \Yb$ as 
\begin{align}
T_{m,n}(x,z) &= (x, z+m+n\tau)~, &
R(x,z) & = (\sigma(x),-z)~.
\end{align}
These obey
\begin{align}
R^2 & = \text{id}~, &
T_{m,n} T_{m',n'} & = T_{m+m',n+n'}~,&
R^{-1} T_{m,n} R & = T_{-m,-n} = (T_{m,n})^{-1}~.
\end{align}
Since the $T^2$ is flat it is clear that loops associated to $N = \Z^2 \subset \pi_1(Y)$ have trivial holonomy.  Thus we have
\begin{equation}
\begin{tikzcd}
1 \ar[r]	& \underbrace{\Hol^0(Y)}_{\simeq \SU(2)} \ar[r]	& \Hol(Y)\ar[r] & \Z_2 \ar[r] &1
\end{tikzcd}~,
\end{equation}
where the $\Z_2$ is generated by $R \in \pi_1(Y)$. Since $Y$ is a Calabi-Yau $3$-fold it must be that $\Hol(Y)$ is also contained in the normalizer subgroup of $\SU(2) \subset \SU(3)$ whose elements are represented by $3\times 3$ matrices
\begin{align}
\cM & = \begin{pmatrix}  A & 0 \\ 0 & \det(A)^{-1} \end{pmatrix}~, &  A \in \GU(2)~.
\end{align}
From our discussion of the Enriques surface it is then clear that $\Hol(Y)$ consists of $\cM$ with $A \in \SU(2)$, as well as an additional element
\begin{align}
\cR & = \begin{pmatrix} \sigma_3 & 0 \\ 0 & -1 \end{pmatrix}~.
\end{align}
So, we conclude that $\Hol(Y) = \SU(2)\rtimes \Z_2$.

\subsection{The unoriented cousin of the Enriques $3$-fold}
Now we consider the unorientable manifold $Z = (\Mb \times T^3) /\kappa_3$, with $\kappa_3 = \sigma \rho_1\rho_2\rho_3$ as in the text.  Its universal cover is $\Zb = \Mb \times \R^3$, and by essentially the same arguments as for the Enriques $3$-fold we have $\pi_1(Z) = \Z^3 \rtimes \Z_2$, $N = \Z^3$, and $\Hol(Z) \subset \GU(2) \times \GO(3) \subset \GO(7)$ is generated by
\begin{align}
\cM & = (A,\iden_3)~, &  \text{with} ~A &\in \SU(2)~,& \text{and}~~
\cR & = (\sigma_3,-\iden_3)~,
\end{align}
so that $\Hol(Z) = \SU(2) \rtimes \Z_2 \subset \GO(7)$.

\subsection{Quotients of $3$-folds}
Finally, we consider $X = \Xb/G$ and $Z = (\Xb \times S^1)/K$, where $\Xb$ is a simply-connected Calabi-Yau $3$-fold, and therefore has $\Hol(\Xb) = \SU(3)$.  In the former case we know that $\Hol(X) = \SU(3)$ as well, so that $N = \pi_1(X)$.   The latter case is more interesting.  Now  $\pi_1(Z) = \Z \rtimes K$, and we can argue that $Q = \Z_2$ as follows.  Recall that the elements of $K$ are graded according to the homomorphism $\phi$ as in~(\ref{eq:Kaction}).  We know that parallel transport along the flat $\R$ direction in the universal cover $\Zb = \Xb \times \R$ has trivial holonomy; in addition, an element of $\pi_1(Z)$ associated to $\kappa \in K$ with $\phi(\kappa) = 0$ can be understood as just a loop in $X$, and therefore must have holonomy valued in $\SU(3)$.  It follows that $Q$ has just one non-trivial element represented by $\varphi( \kappa)$, where $\kappa \in \pi_1(Z)$ acts on $\Zb$ as $\kappa(x,\theta) = (\sigma(x),-\theta)$, and there is, once again, a short exact sequence
\begin{equation}
\begin{tikzcd}
1 \ar[r]	& \underbrace{\Hol^0(Z)}_{\simeq \SU(3)} \ar[r]	& \Hol(Z)\ar[r] & \Z_2 \ar[r] &1
\end{tikzcd}~.
\end{equation}
The normalizer subgroup of $\SU(3)$ inside $\GO(7)$ is $\GU(3)\times \Z_2$, and the element $\cT \in \Hol(Z)$ that maps to the generator of $\Z_2$ has the form
\begin{align}
\cT &= (\cM,-1)~,\qquad \cM \in \GU(3)~,
\end{align}
where $\cM$ is in the fundamental representation of $\GU(3)$ and therefore can be written as
\begin{align}
\cM &= e^{\ii\psi} \cS~, &  \cS \in \SU(3)~.
\end{align}
The pair $(\psi,\cS)$ is ambiguous:  $(\psi,\cS) \sim (\psi + 2\pi/3, \cZ \cS)$, where $\cZ$ generates the center of $\SU(3)$, and since it must also be that $\cM^2 \in \SU(3)$, we can restrict attention to two possibilities:  $\psi = 0$ or $\psi = \pi/3$.  The second possibility is inconsistent because it would imply that there is a loop $\gamma: S^1\to X$ with $\cP_{\gamma} \not\in\SU(3)$.   So, it must be that $\cM \in \SU(3)$.  $\Hol(Z)$ is generated by elements $(\cX,1)$, with $\cX \in \SU(3)$, as well as $\cT = (\cM,-1)$.  Because $\cM\in \SU(3)$ we can instead replace $\cT$ by $\cT' = (\cM^\dag,1) \cdot \cT = (\iden_3,-1)$, which shows that $\Hol(Z) = \SU(3)\times \Z_2$.   This result should be contrasted with the Enriques examples discussed above, where the analog of $\cM \in \SU(3)$ was instead $\sigma_3 \in \GU(2)$, leading to the non-trivial semi-direct product in the full holonomy group.

\section{T-duality, worldsheet parity, and fermion numbers}\label{sec:T}
In this section we consider type II string theory on $T^d$ and recall the consequences of lifting the familiar operations of T-duality, denoted by $\cT_i$ for the $i$-th circle, reflections $\cR_i$, and worldsheet parity $\Pi$, initially defined in the NS-NS sector,  to the Ramond sectors. 

If we define, in the NS-NS sector, $\cT_i$ to be a reflection of the right-moving fields associated to the $i$-th circle, then we can choose the lift of $\cT_i$ to the Ramond sectors so that the following relations hold:
\begin{align}
\cT_i^2 & = (-1)^{\bF_{\sright}}~,&
\cR_i^2 & = (-1)^{\bF}~,&
\cT_i \Pi \cT_i^{-1}  & = \cR_i \Pi~,&
\cT_i \cR_i \cT_i^{-1} & = \cR_i~, &
\Pi \cR_i & = \cR_i \Pi (-1)^{\bF}~,
\end{align}
and also for $i\neq j$
\begin{align}
\cT_i \cR_j \cT_i^{-1} & = (-1)^{\bF_{\sright}} \cR_j~, &
\cR_i \cR_j & = (-1)^{\bF} \cR_j \cR_i~.
\end{align}
These can be easily worked out using free fields.  Denoting the linear combinations of the R--NS and NS--R ground states by the column vector,
\begin{align}
\Sigma & = \begin{pmatrix}
S_{\sleft} \\ S_{\sright}
\end{pmatrix}~,
\end{align}
we choose the following lifts:
\begin{align}
(-1)^{\bF_{\sleft}} \cdot \Sigma & = \begin{pmatrix}
-\iden & 0 \\ 0 & \iden 
\end{pmatrix} \Sigma~,&
(-1)^{\bF_{\sright}} \cdot \Sigma & = \begin{pmatrix}
\iden & 0 \\ 0 & -\iden 
\end{pmatrix} \Sigma&
\Pi \cdot \Sigma & = \begin{pmatrix}
0 & \iden \\ \iden & 0 
\end{pmatrix}\Sigma~, \nonumber\\
\cT_i \cdot \Sigma & =  \begin{pmatrix}
 \iden & 0 \\ 0 & \Gamma_i \Gamma
\end{pmatrix} \Sigma~,&
\cT_i^{-1} \cdot \Sigma & =  \begin{pmatrix}
 \iden & 0 \\ 0 &- \Gamma_i \Gamma
\end{pmatrix} \Sigma~,&
\cR_i \Sigma & = \begin{pmatrix} - \Gamma_i \Gamma & 0 \\ 0 & \Gamma_i \Gamma\end{pmatrix} \Sigma~,
\end{align}
and the relations quoted above follow.

\bibliographystyle{./utphys}
\bibliography{./newref}

\end{document}